
\input phyzzx
\input epsf
\FRONTPAGE
\line{\hfill BROWN-HET-995}
\line{\hfill April 1995}
\bigskip
\titlestyle{ACCRETION OF HOT DARK MATTER ONTO SLOWLY MOVING COSMIC STRINGS}
\medskip
\author{Anthony N. Aguirre$^a$ and Robert H. Brandenberger$^{a,b}$}
\centerline{{\it a) Physics Department}}
\centerline{{\it Brown University, Providence, RI 02912, USA}}
\smallskip
\centerline{\it b) Physics Department}
\centerline{\it University of British Columbia}
\centerline{\it Vancouver, B.C. V6T 1Z1, CANADA}
\medskip
\abstract
Cosmic strings with small-scale structure have a coarse-grained mass per unit
length $\mu$ which is larger than the string tension. This leads to an
effective Newtonian gravitational line source and to a characteristic
translational velocity which is smaller than for strings without small-scale
structure. Here, the accretion of hot dark matter onto such strings is studied
by means of the Zel'dovich approximation. We find that clustering is greatly
enhanced by the Newtonian line source. In the limit of vanishing translational
velocity, the first nonlinear filaments form at a redshift of greater than 100
for standard values of $\mu$.
\endpage
\chapter{Introduction}

A theory in which the dark matter is hot and in which the primordial density
perturbations are seeded by cosmic strings is a viable alternative$^{1,2)}$ to
the currently popular variants of cold dark matter (CDM) cosmologies with
adiabatic density fluctuations produced during a hypothetical period of
inflation in the very early Universe. On large length scales, the theory
predicts a scale-invariant spectrum of perturbations. The normalizations of the
model from large-scale structure$^{3)}$ and from cosmic microwave background
anisotropy measurements$^{4)}$ are in encouraging agreement. The theory also
predicts that on scales comparable to and larger than the comoving horizon at
$t_{eq}$, the time of equal matter and radiation, galaxies are concentrated in
sheets or filaments$^{3,5-8)}$, in agreement with the results from the Center
for Astrophysics galaxy redshift survey$^{9)}$.

If the primordial density inhomogeneities are nonadiabatic, slowly decaying
seeds such as cosmic strings, hot dark matter (HDM) is a viable dark matter
candidate. In contrast to adiabatic inflationary models in which free streaming
erases perturbations on all scales smaller than the maximal free streaming
length $\lambda_J^{max}$, the strings survive neutrino free streaming$^{10)}$
and can at late times seed structures on scales much smaller than
$\lambda_J^{max}$. There is$^{11)}$ still power on scales smaller than
$\lambda_J^{max}$ (although less than in the inflation-based CDM cosmology),
and the power spectrum may well be in better agreement with the results from
recent observations$^{12)}$ (the decrease in power on small scales compared to
the inflationary CDM model is a promising feature).

In spite of the abovementioned advantages of the cosmic string theory, there is
still a lot of ignorance about the specific predictions of the model. One
important reason is that the initial conditions for structure formation in the
theory are not yet accurately known. It is well established that the network of
cosmic strings (and thus the distribution of the seeds for density
inhomogeneities) approaches a scale-invariant distribution, but the detailed
properties of this distribution are uncertain. It can be shown that at all
times $t$, there are a few strings which are straight on a correlation length
$\xi (t) \sim t$ crossing each Hubble volume, but the small-scale structure of
these strings is not known. Also uncertain is the exact number of such strings
per Hubble volume.

The evolution of the cosmic string network may leave behind on the strings a
lot of structure (kinks and wiggles) on scales smaller than $\xi$. The typical
amplitude of these perturbations is much smaller than $\xi$, and hence in a
coarse grained description, the strings can still be viewed as straight on a
scale $\xi$, but with renormalized mass per unit length and tension. Since the
integrated length of the string between two points is longer than the
separation between the points, the coarse grained mass per unit length is
larger than the microscopic length. At the same time, the coarse grained
tension is less than the microscopic tension. As first discussed by
Carter$^{13)}$ and subsequently analyzed in detail by Vollick$^{6)}$ and by
Vachaspati and Vilenkin$^{7)}$, small-scale structure on strings can lead to
substantial modifications of the motion and gravitational effects of the
strings.

A long straight string without small-scale structure induces no local
gravitational force. However, space perpendicular to the string is a cone with
deficit angle$^{14)}$
$$
\alpha = 8 \pi G \mu , \eqno\eq
$$
where $\mu$ is the mass per unit length of the string coarse grained on a scale
$\xi$ and $G$ is Newton's constant. When moving with a transverse speed $v_s$,
such a string will induce a velocity perturbation of magnitude
$$
\delta v = 4 \pi G \mu v_s \gamma_s \eqno\eq
$$
towards the plane behind the string, leading to the formation of a ``wake", a
planar density enhancement$^{5)}$. In the above, $\gamma_s$ is the relativistic
$\gamma$ factor associated with $v_s$.

A long straight string with small-scale structure will induce, in addition to
the deficit angle given by Eq. (1.1), a nonvanishing local gravitational force
of magnitude$^{6,7)}$
$$
F = 2 m G (\mu - T) / r \eqno\eq
$$
towards the string when acting on a test particle of mass $m$ located a
distance $r$ from the string. The string tension is $T$, and $T$ is related to
$\mu$ by$^{13)}$
$$
\mu T = \mu_0^2, \eqno\eq
$$
where $\mu_0$ is the microscopic mass per unit length of the string.
In addition, when coarse grained on a scale of $\xi$, strings with
$\mu > \mu_0$ have a smaller transverse velocity than those with $\mu = \mu_0$.
In the limit when the distance the string moves in one Hubble expansion time is
small compared to the thickness of the wake, the structure seeded by the string
becomes filamentary.

It is important to know at what redshift the first density perturbations become
nonlinear (this gives an upper bound to the redshift of first star formation).
To answer this question in our theory, the clustering of HDM induced by cosmic
strings must be studied.

The accretion of HDM onto cosmic string loops was studied in detail in Ref. 10.
It was shown that free streaming slows down but never reverses the growth of
perturbations on small scales. Structures form ``inside-out" in the sense that
the first shells to go nonlinear around the seed are the innermost ones. In a
toy model in which only cosmic string loops are present as seeds, this leads to
a ``bottom-up" scenario for structure formation. The first nonlinear objects
form at high redshifts.

However, the current cosmic string simulations$^{15)}$ indicate that most of
the mass in the string network resides in the long strings (strings with
curvature radius $\xi(t) \sim t$ at cosmic time $t$). Hence, in order to
determine the earliest time at which large-scale nonlinear structures can form,
it is necessary to study the clustering of HDM induced by long strings. This
was done in Ref. 3 for strings without small-scale structure. The results were
very different than for loops. Planar collapse of HDM does not proceed in an
``inside-out" fashion. Rather, the first sheet to go nonlinear has an initial
comoving distance $q_{max} \sim 1$ Mpc from the center of the wake, and the
onset of nonlinearity occurs late, i.e. at a redshift $z_{max} \sim 1$ for
values $G \mu \sim 10^{-6}$ (these results will be reviewed at the end of
Section 2).

In this paper, we study the accretion of HDM onto long strings with
nonvanishing small-scale structure (which leads to the presence of a Newtonian
gravitational line source). The calculations will be done in the limit in which
the tranlational motion of the string can be neglected.

We find that in this situation, structures once again form ``inside-out", and
that the first nonlinear structures emerge at a high redshift ($z_{max} > 10^2$
for $G \mu \sim 10^{-6}$). This result indicates that in a cosmic string theory
with HDM, there will be no problem in explaining the recent observations of
high redshift clusters$^{16)}$, quasars$^{17)}$ and QSO absorption line
systems$^{18)}$.

In the following section we summarize our calculations, and in Section 3 we
discuss some implications of our results. Units in which $c = k_B = 1$ are used
throughout. The scale factor of the Universe is denoted by $a(t)$, $t$ being
cosmic time, and the associated redshift is $z(t)$. The present time is $t_0$.

\chapter{Accretion of Hot Dark Matter onto a Line Source}

We will study the accretion of collisionless hot dark matter (HDM) onto a
stationary cosmic string using an adaptation of the Zel'dovich
approximation$^{19)}$ to HDM. This modification was applied to planar collapse
in Ref. 3 and was shown to give good agreement with the results of a linearized
Boltzmann equation method (which takes into account the full phase space
distribution of HDM particles; see Refs. 3 \& 10).

HDM particles have near-relativistic velocities $v(t)$ at times close to
$t_{eq}$. This prevents them from clustering on length scales smaller than
their free streaming distance
$$
\lambda_J(t) \simeq 3 v(t) z(t) t, \eqno\eq
$$
the average comoving distance a particle will move in a Hubble expansion time.
Note that since $v(t) \propto z(t) \propto a(t)^{-1}$, the free streaming
length decreases in time as
$$
\lambda_J(t) \propto t^{-1/3} \eqno\eq
$$
and takes on its maximal value $\lambda_J^{max}$ at $t_{eq}$. In a spatially
flat HDM Universe
$$
v(t_{eq}) = v_{eq} \simeq 0.1 \eqno\eq
$$
and hence
$$
\lambda_J^{max} \simeq 3 h^{-2} Mpc, \eqno\eq
$$
substantially larger than the typical size of a galaxy.

     We will first describe the Zel'dovich approximation for filamentary
accretion and then in a second step adapt the method to HDM. Without loss of
generality we can choose coordinates in which the string lies along the $z$
coordinate axis. We consider the response of the dark matter particles to the
gravitational line source produced by the small-scale structure of the string,
focusing on the ``plane" (strictly speaking a cone) at $z = 0$. Let $\vec q$
denote the initial comoving separation of a dark matter particle from the
string. In the absence of the string line source, the physical separation $\vec
r(t)$ would grow as the Universe expands. The presence of the string, however,
leads to a comoving displacement $\vec \psi(\vec q , t)$ which grows in time:
$$
\vec r(\vec q,t)=a(t)(\vec q - \vec \psi(\vec q,t)). \eqno\eq
$$

     The Zel'dovich approximation consists of taking the Newtonian equations
$$
\ddot{\vec r} = - {{\partial} \over {\partial \vec r}} \Phi (\vec r,t) \eqno\eq
$$
for $\vec r(\vec q , t)$ with the gravitational potential $\Phi(\vec r , t)$
determined by the Poisson equation
$$
{{\partial ^2} \over {\partial \vec r ^2}} \Phi = 4\pi G\left(\rho(\vec r,t)+
{{\lambda \delta (r)} \over {2\pi r}}\right) \eqno\eq
$$
and linearizing in $\psi$. On the right hand side of the Poisson equation, the
total density has been separated into the dark matter density $\rho(\vec r ,
t)$ and the contribution of the string-induced line source with an effective
mass per unit length
$$
\lambda = \mu - T \eqno\eq
$$
(see Eq. (1.3)), and where $\delta(r)$ denotes a Dirac delta function in
cylindrical coordinates. The density $\rho(\vec r , t)$ is in turn determined
in terms of the background density $\rho_0(t)$ by the mass conservation
equation
$$
\rho (\vec r,t)d^3\vec r= a^3 (t)\rho _0 d^3\vec q \eqno\eq
$$
which gives
$$
\rho (\vec r,t) = a^3 (t) \rho_0 det^{-1}\left| {{d\vec r} \over {d\vec
q}}\right|
= a(t) \rho_0 ({{dr} \over {dq}})^{-1} \simeq \rho_0 (1 + {{\partial} \over
{\partial q}} \psi_r(q, t)), \eqno\eq
$$
where $r$, $q$ and $\psi_r$ are (in cylindrical coordinates) the radial
components of $\vec r$, $\vec q$ and $\vec \psi$, respectively, and where in
the final step we have linearized in $\vec \psi$.

Inserting (2.10) into (2.7) and integrating once gives
$$
{{\partial} \over {\partial \vec r}}\Phi = 4\pi G\left[{{1} \over {3}}\rho
_0(t)\vec r + \rho _0(t)a(t)\vec \psi(\vec q,t) + {{\lambda\theta(r)\hat r}
\over {2\pi r}}\right],
\eqno\eq
$$
which we insert on the right hand side of (2.6). Note that $\hat r$ is the unit
radial vector in the plane perdendicular to the string. On the left hand side
of (2.6) we substitute the second time derivative of (2.5),
$$
-\ddot{\vec r} = a\ddot{\vec \psi} + 2\dot a \dot{\vec \psi} - \ddot a
\dot{\vec q} + \ddot a \vec \psi. \eqno\eq
$$
Using the FRW equation $\ddot a = - {{4 \pi G} \over 3} \rho_0 a$ (valid in the
matter dominated era) to substitute for $\rho_0 a$ in (2.11), we obtain as our
basic equation
$$
\ddot{\psi_r} + 2 {{\dot a} \over {a}}\dot{\psi_r} + 3{{\ddot a} \over {a}}
\psi_r = {{2\lambda G} \over {a^2(q -  \psi_r)}}. \eqno\eq
$$
This equation is only valid for $t > t_{eq}$, since we made use of an
approximate pressureless equation of state. Hence, we should also substitute
$a(t) = (t/t_0)^{2/3}$ and then obtain
$$\ddot \psi_r + {{4} \over {3}}t^{-1}\dot \psi_r-{{2} \over {3}}t^{-2}\psi_r =
{{2\lambda G t_0^{4/3}} \over {t^{4/3}(q - \psi_r)}} \eqno\eq
$$
as equation of motion for the radial displacement $\psi_r$.

A second reason why the above derivation is only valid for $t > t_{eq}$ is that
the presence of the uniform radiation bath has been neglected. This radiation
would have to be considered in the Poisson equation (2.7) and would stall the
growth of perturbations before $t_{eq}$. Hence, for a cosmic string theory with
cold dark matter, the initial conditions for solving (2.14) are
$$
\dot{\vec \psi}(t_s) = \vec \psi(t_s) = 0, \eqno\eq
$$
with $t_s = t_{eq}$ (for strings present at $t_{eq}$) and $t_s = t_i$ for
strings ``appearing" at a later time.

For a static string network, it does not make sense to consider $t_i > t_{eq}$.
However, for a dynamical string network, $t_i > t_{eq}$ would apply to string
segments whose curvature radius $\xi$ is equal to the Hubble radius at time
$t_i$ and thus start their proper motion at that time.

To first order in perturbation theory, we solve (2.14) using the Born
approximation
$$
\ddot \psi_r + {{4} \over {3}}t^{-1}\dot \psi_r - {{2} \over {3}}t^{-2}\psi_r =
{{2\lambda t_o^{4/3} G} \over {q t^{4/3}}}. \eqno\eq
$$
This can be solved using the Green's function method which gives
$$
\psi_r(t) = u_1(t)\int_{t_s}^{t}{dt'\epsilon(t')u_2(t')f(t')} -
u_2(t)\int_{t_s}^{t} {dt'\epsilon(t')u_1(t')f(t')}, \eqno\eq
$$
where $u_1(t) = ({t \over {t_0}})^{2/3}$ and $u_2(t) = ({t \over {t_0}})^{-1}$
are the two fundamental solutions of the homogeneous equation, $\epsilon(t)$ is
the Wronskian
$$
\epsilon(t) = (\dot u_1 u_2 - \dot u_2 u_1)^{-1}, \eqno\eq
$$
and
$$
f(t) = {{2 \lambda G} \over {q}} \left( {{t_0} \over {t}}\right) ^{4/3}.
\eqno\eq
$$
Some manipulations give the end result:
$$
\psi_r(t) = {{6\lambda G} \over {5q}}t_0^{4/3}t^{2/3}\left[ln(t/ts) -
{3 \over 5}\left(1-\left({{t_s} \over {t}}\right)^{5/3}\right)\right].\eqno\eq
$$

The real objective of the calculation, however, is to find the comoving
scale $q_{nl}$ which has gone non-linear at a given time $t$, as a function
of the time of formation $t_i$.  This scale is defined as the scale at which
at a time $t$ the physical velocity of a particle $\dot r(q_{nl},t)$ vanishes,
giving an equation for $q_{nl}$ of
$$
\dot a(t)(q_{nl}(t_s,t)-\psi(t)) = a(t)\dot \psi(t). \eqno\eq
$$
Together with Eq. (2.20) and the assumption that $a(t) = (t/t_0)^{2/3}$,
this gives (after a fair amount of algebra)
$$
q_{nl}(t_s,t) = t^{1/3}t_0^{2/3}\sqrt{{{12G\lambda} \over {5}}\left[
log(t/t_s) - {3 \over {20}}\left( \left( {{t_s} \over {t}}\right) ^{5/3} -
1\right) \right] }. \eqno\eq
$$
For CDM, this equation gives the size of the non-linear region at time $t$ for
a structure which starts to grow at time $t_s$.

Because of neutrino free streaming, the above formalism is not directly
applicable to the clustering of HDM. A rigorous analysis must follow the entire
phase space distribution of the dark matter. Starting point is the
collisionless Boltzmann equation describing the time evolution of the phase
space density, expanded perturbatively about a homogeneous thermal
distribution. After integrating over momenta one obtains the Gilbert
equation$^{20)}$, an integral equation for the configuration space density
perturbation. However, in Ref. 3 it was shown that the Zel'dovich approximation
can easily be adapted to HDM and then gives good agreement with the results
from the Gilbert equation approach.

The modified Zel'dovich approximation consists of using the same dynamical
equation (2.14), but with initial conditions
$$
\dot{\vec \psi}(t_s(q)) = \vec \psi(t_s(q)) = 0, \eqno\eq
$$
where $t_s(q)$ is the time when the neutrino free streaming length
$\lambda_J(t)$ equals $q$, i.e.
$$
\lambda_J(t_s(q)) = q . \eqno\eq
$$
Solving for $t_s(q)$ gives
$$
t_s = {{t_0\lambda _0^3} \over {q^3}}, \ \ \ \ \ \lambda _0 = 3 v_{eq}
z_{eq}^{1/2} t_{eq}. \eqno\eq
$$
Note that if (2.24) gives a value of $t$ smaller than $t_i$, then $t_s(q)$ must
again be chosen to be $t_i$.

For HDM, turnaround in determined by Equation (2.22), but with the time $t_s$
depending on $q$ as given by (2.25). Inserting (2.25) into (2.22) and solving
for $q$ at late times gives
$$
{{6G\lambda t_0^2} \over {5z(t)}}\left(6 log\left({{q} \over {z^{1/2}(t)\lambda
_0}} \right) + {{3} \over {10}}\right) - q^2 = 0. \eqno\eq
$$
This implicit equation gives, for a given scale $q$, the redshift $z(t)$ at
which this scale $q$ first goes non-linear. This equation only holds for values
of $q$ smaller than $q_{max}$, the value of $q$ for which $t_s(q)$ equals
$t_i$, i.e.
$$
q_{max} = \lambda_0 z^{1/2}(t_i) . \eqno\eq
$$
For $q$ larger than $q_{max}$, Equation (2.25) with $t_s = t_i$ must be used.

Equation (2.26) can be solved numerically. The result is shown in Figure 1. The
main conclusion is that in the filament model with HDM, non-linear regions can
develop at very early times - at redshifts larger than $10^2$ for the usual
cosmic string normalization $G \mu = 10^{-6}$. The first scales to go
non-linear are small scales.

These results contrast with those for planar collapse of HDM$^{3)}$, where no
non-linearities appear until a redshift of about 1, and the first scale to turn
around is $q = q_{max}$ (see Figure 2).
\endpage
\vskip.5in
\epsfxsize=4in \epsfbox{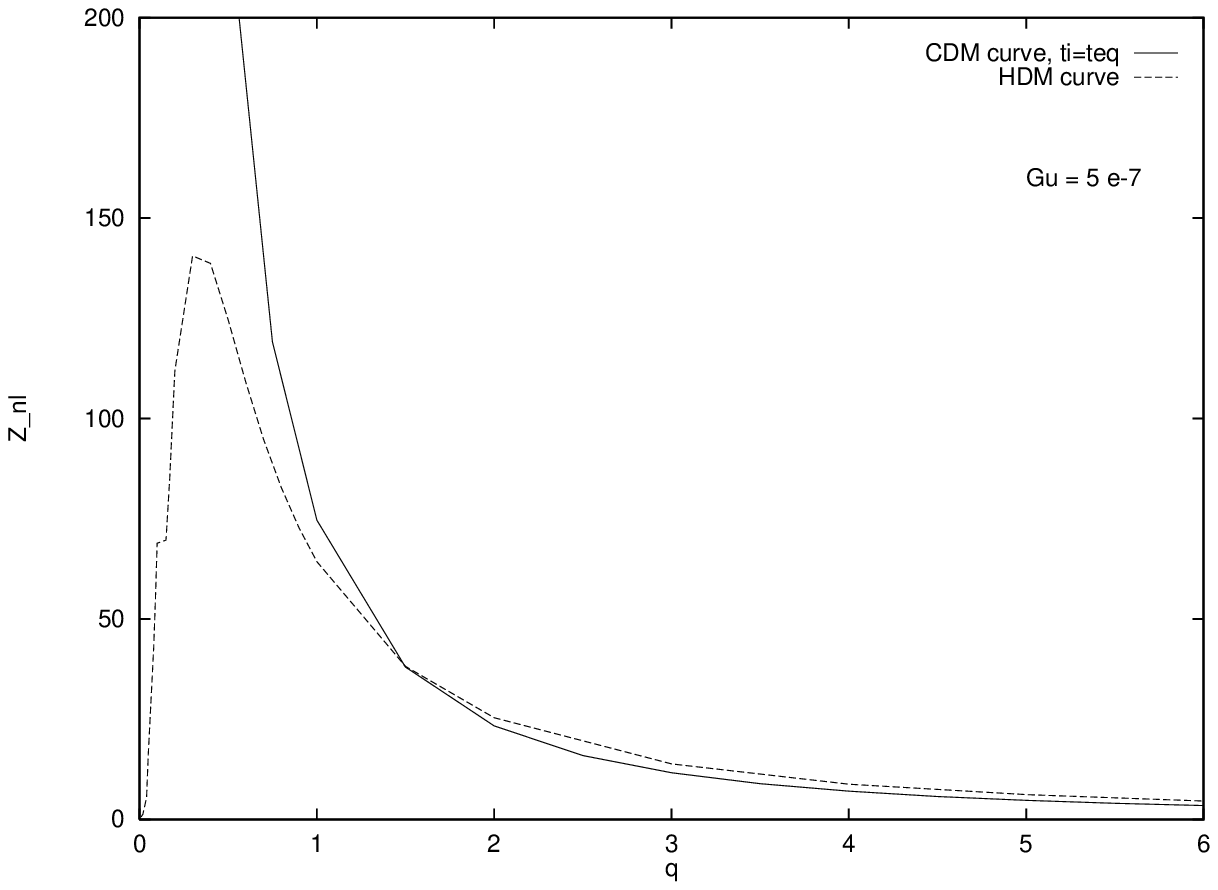}
{\baselineskip=13pt
\noindent{\bf Figure 1:} Redshift $z_{nl}$ at which a scale $q$ (in Mpc) turns
around for filamentary accretion in the case of HDM (dashed line) and CDM
(solid line). The value of $G \lambda$ was $5 \cdot 10^{-7}$, and the case $t_i
= t_{eq}$ was considered.}

\vskip.5in
\epsfxsize=4in \epsfbox{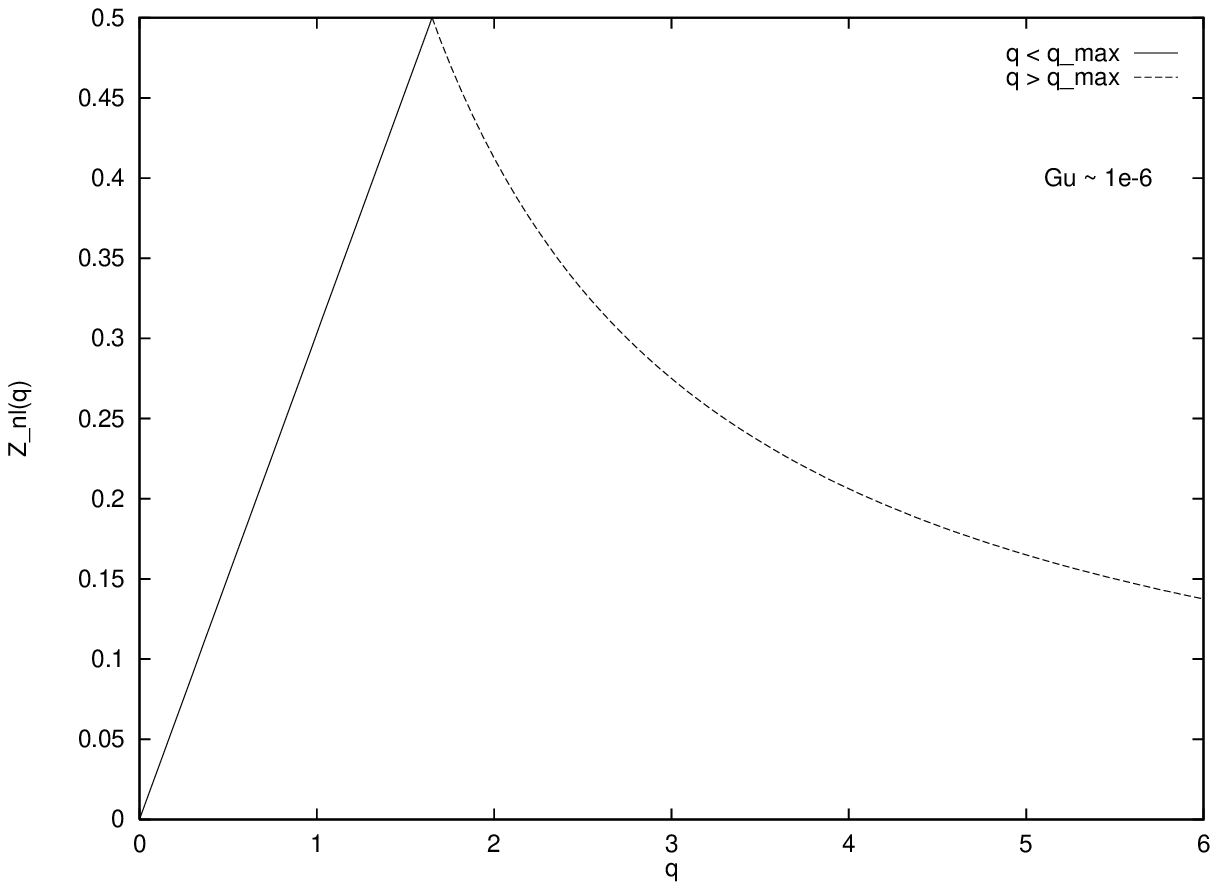}
{\baselineskip=13pt
\noindent{\bf Figure 2:} Redshift $z_{nl}$ versus scale $q$ (in Mpc) for
accretion of HDM onto a cosmic string wake for a value $G \mu = 10^{-6}$ (from
Ref. 3).}

\chapter{Discussion}

We have studied the accretion of HDM onto a stationary cosmic string with
nonnegligible small-scale structure. Such a string acts as a Newtonian line
source of gravity for HDM. Our main results are summarized in Figure 1: the
first scales to turn around are small scales, and the redshift when the first
non-linearities appear exceeds $10^2$ for the usual cosmic string
normalization. This contrasts with the results of planar collapse of HDM onto
cosmic string-induced wakes, in which case no non-linear structures form until
a reshift of about 1, and where large scales $q = q_{max}$ are the first to
turn around.

These results are important in determining the redshift when the first
non-linear filamentary structures form in a cosmic string model with HDM. The
mass of a filament seeded at $t_{eq}$ by a horizon-length string is
$$
M(z) \sim \rho_0 t_{eq} z(t_{eq}) q_{nl}^2(z) \sim 10^{12} M_O ( {{q_{nl}(z)}
\over {Mpc}})^2, \eqno\eq
$$
where $\rho_0$ is the background energy density and $M_O$ is a solar mass. Such
a mass is large enough to be the host of a quasar at a redshift much greater
than 1$^{21)}$. Hence, explaining the origin of high redshift quasars and
galaxies seems not to be a problem in this model.

Obviously, much more work is needed in order to be able to reliably estimate
the high redshift mass function in the cosmic string model with HDM. Besides
the filamentary accretion mechanism discussed in this paper, there is also
accretion at high redshift onto cosmic string loops$^{10)}$. Which of the
mechanisms is dominant unfortunately depends very sensitively on the parameters
in the cosmic string scaling solutions which are not yet reliably known.

Considering a long cosmic string as stationary in order to discuss the
accretion of HDM on a given comoving scale $q$ will only be a reasonable
approximation if $q$ is larger than the comoving distance the string moves at
$t_i$ in one Hubble expansion time, i.e.
$$
q > t_0 z(t_i)^{-1/2} v_s, \eqno\eq
$$
where $v_s$ is the transverse velocity of the string. For $t_i = t_{eq}$, this
gives
$$
q > 30 v_s Mpc. \eqno\eq
$$
For strings with $v_s \propto 0.03$, the approximation of filamentary accretion
is hence reasonable for scales $q \propto 1 Mpc$. Note that the r.m.s. value of
$v_s$ is about $0.15$ in the current cosmic string simulations$^{15)}$ which
probably underestimate the effects of small-scale structure and thus
overestimate $v_s$.

In conclusion, we believe our results are good news for the cosmic string model
with HDM and should encourage further research. The first non-linear structures
form early enough in order to explain the origin of high redshift quasars,
galaxies and clusters. The results also imply that the value of $G \mu$ could
be lowered from its canonical value of $G \mu \simeq 10^{-6}$ without
jeopardizing early structure formation.
\bigskip
\centerline{\bf Acknowledgements}

This work is supported in part by the US Department of Energy under
Grant DE-FG0291ER40688, Task A (Brown) and by the Canadian NSERC under
Grant 580441 (UBC). One of us (R.B.) is grateful to Richhild Moessner for
useful conversations, and to Bill Unruh for his warm hospitality at UBC, where
this article was written up. This work is based on the senior thesis of A.A. at
Brown University.

\REF\one{R. Brandenberger, {\it Phys. Scripta} {\bf T36}, 114 (1991).}
\REF\two{A. Vilenkin and E.P.S. Shellard, {\it Cosmic strings and other
topological defects} (Cambridge Univ. Press, Cambridge, 1994); \nextline
M. Hindmarsh and T.W.B. Kibble, {\it Cosmic strings}, subm. to Rep. Prog.
Phys., hep-ph/9411342; \nextline
R. Brandenberger, {\it Int. J. Mod. Phys.} {\bf A9}, 2117 (1994).}
\REF\three{L. Perivolaropoulos, R. Brandenberger and A. Stebbins,, {\it Phys.
Rev.} {\bf D41}, 1764 (1990); \nextline
R. Brandenberger, L. Perivolaropoulos and A. Stebbins, {\it Int. J. Mod. Phys.}
{\bf A5}, 1633 (1990).}
\REF\four{D. Bennett, A. Stebbins and F. Bouchet, {\it Ap. J. (Lett.)} {\bf
399}, L5 (1992); \nextline
L. Perivolaropoulos, {\it Phys. Lett.} {\bf B298}, 305 (1993).}
\REF\five{Ya. B. Zel'dovich, {\it Mon. Not. R. astr. Soc.} {\bf 192}, 663
(1980); \nextline
J. Silk and A. Vilenkin, {\it Phys. Rev. Lett.} {\bf 53}, 1700 (1984);
\nextline
T. Vachaspati, {\it Phys. Rev. Lett.} {\bf 57}, 1655 (1986); \nextline
A. Stebbins, S. Veeraraghavan, R. Brandenberger, J. Silk and N. Turok, {\it Ap.
J.} {\bf 322}, 1 (1987).}
\REF\six{D. Vollick, {\it Phys. Rev.} {\bf D45}, 1884 (1992); \nextline
D. Vollick, {\it Ap. J.} {\bf 397}, 14 (1992).}
\REF\seven{T. Vachaspati and A. Vilenkin, {\it Phys. Rev. Lett.} {\bf 67}, 1057
(1991); \nextline
T. Vachaspati, {\it Phys. Rev.} {\bf D45}, 3487 (1992).}
\REF\eight{T. Hara and S. Miyoshi, {\it Prog. Theor. Phys.} {\bf 77}, 1152
(1987); \nextline
T. Hara and S. Miyoshi, {\it Prog. Theor. Phys.} {\bf 84}, 867 (1990).}
\REF\nine{V. de Lapparent, M. Geller and J. Huchra, {\it Ap. J. (Lett.)} {\bf
302}, L1 (1986); \nextline
M. Geller and J. Huchra, {\it Science} {\bf 246}, 897 (1989).}
\REF\ten{R. Brandenberger, N. Kaiser, D. Schramm and N. Turok, {\it Phys. Rev.
Lett.} {\bf 59}, 2371 (1987); \nextline
R. Brandenberger, N. Kaiser and N. Turok, {\it Phys. Rev.} {\bf D36}, 2242
(1987); \nextline
E. Bertschinger and P. Watts, {\it Ap. J.} {\bf 328}, 23 (1988).}
\REF\eleven{A. Albrecht and A. Stebbins, {\it Phys. Rev. Lett.} {\bf 69}, 2615
(1992).}
\REF\twelve{see e.g. K. Fisher et al., {\it Ap. J.} {\bf 402}, 42 (1993).}
\REF\thirteen{B. Carter, {\it Phys. Rev.} {\bf D41}, 3869 (1990).}
\REF\fourteen{A. Vilenkin, {\it Phys. Rev.} {\bf D23}, 852 (1981);
\nextline
J. Gott, {\it Ap. J.} {\bf 288}, 422 (1985); \nextline
W. Hiscock, {\it Phys. Rev.} {\bf D31}, 3288 (1985); \nextline
B. Linet, {\it Gen. Rel. Grav.} {\bf 17}, 1109 (1985); \nextline
D. Garfinkle, {\it Phys. Rev.} {\bf D32}, 1323 (1985); \nextline
R. Gregory, {\it Phys. Rev. Lett.} {\bf 59}, 740 (1987).}
\REF\fifteen{D. Bennett and F. Bouchet, {\it Phys. Rev. Lett.} {\bf 60}, 257
(1988);\nextline
B. Allen and E.P.S. Shellard, {\it Phys. Rev. Lett.} {\bf 64}, 119
(1990);\nextline
A. Albrecht and N. Turok, {\it Phys. Rev.} {\bf D40}, 973 (1989).}
\REF\sixteen{F. Castander, R. Ellis, C. Frenk, A. Dressler and J. Gunn, {\it
Ap. J. (Lett.)} {\bf 424}, L79 (1994);\nextline
R. Nichol, M. Ulmer, R. Kron, G. Wirth and D. Koo, {\it Ap. J.} {\bf 432}, 464
(1994);\nextline
G. Luppino and I. Gioia, Univ. of Hawai preprint, astro-ph/9502095, {\it Ap. J.
(Lett.)}, in press (1995).}
\REF\seventeen{S. Warren, P. Hewett and P. Osmer, {\it Ap. J. (Suppl.)} {\bf
76}, 23 (1991);\nextline
M. Irwin, J. McMahon and S. Hazard, in {\it The Space Distribution of Quasars},
ed. D. Crampton (ASP, San Francisco, 1991), p. 117;\nextline
M. Schmidt, D. Schneider and J. Gunn, ibid., p. 109;\nextline
S. Warren, P. Hewett and P. Osmer, {\it Ap. J.} {\bf 421}, 412 (1994).}
\REF\eighteen{G. Kauffmann and S. Charlot, {\it Ap. J. (Lett.)} {\bf 430}, L97
(1994);\nextline
H. Mo and J. Miracle-Escud\'e, {\it Ap. J. (Lett.)} {\bf 430}, L25 (1994).}
\REF\nineteen{Ya. B. Zel'dovich, {\it Astr. Astrophys.} {\bf 5}, 84 (1970).}
\REF\twenty{J. Bond and A. Szalay, {\it Mon. Not. R. astr. Soc.} {\bf 222}, 27
(1984).}
\REF\twentyone{see e.g. T. Padmanabhan, {\it Structure formation in the
universe} (Cambridge Univ. Press, Cambridge 1993), ch. 9.}
\refout
\bye